\documentstyle[12pt]{article}

\begin{document}

\title{\bf More Insight into Heavy Quark Masses from QCD }

\author{{\bf M. Chabab$^{(a,b)}$\thanks{e-mail:ucadp@cybernet.net.ma} ,
 R. Markazi$^{(a)}$, E.H. Saidi$^{(a)}$}\\
 \\
 \and
 \normalsize{$^{a)}$UFR-High Energy Physics, Physics Department, Faculty of Science}\\
 \normalsize{PO Box 1014, Rabat-Morocco.}\\
 \and
\normalsize{${b)}$ HEP and Astrophysics Group, Physics Department, Faculty of Science}\\
\normalsize{ -Semlalia, PO Box S15, Marrakech 40 000, Morocco.}}

\date{\small}


\maketitle

\begin{abstract}
Using the non-relativistic version of Borel sum rules, combined to the very
accurate experimental information on the Charmonium and Upsilon spectra, we
rigorously investigate the Euclidean mass of the Charm and Bottom quarks.
Our analysis is performed with an improved expansion of the vector
correlator function in the infinite heavy quark mass limit. The optimal estimates,
which take into account the most recent world average value of the strong coupling
constant, as well as a conservative range of the gluon condensate values, are:
$m_c^{Eucl}=1.20\pm 0.034GeV$,$m_b^{Eucl}=4.18\pm 0.037GeV$. Their
conversions to the corresponding pole mass and running mass of c and b quaks
 give respectively: $M_c=1.49\pm 0.08 GeV$, $M_b=4.65\pm 0.06 GeV$ and
${m}_c^{}(\overline{m}_c)=1.21\pm 0.08 GeV$, ${m}_b^{}(\overline{m}_{b)}=4.20\pm 0.06 GeV$
in good agreement with the most recent direct estimates.

\end{abstract}

\vskip -18.5truecm
\rightline{\small UFR-HEP/98-05}
\rightline{\small GHEPA/98-02}
\rightline{\small Mai 1998}
\newpage

\section{Introduction:}

Heavy quark masses play a central role in the Standard Model of particle
physics. They are often present for the phenomenology of a plethora of
processes(weak decays, charm or bottom production,...). However, they cannot
be estimated within the theory . In addition, because of their confinement
at large scales, they are invisible to the experimentalists. In this
context, different theoretical analyses have been used to the extraction of
the charm and bottom quark masses. Besides lattice calculations, QCD
sum-rules -\`{a}-la Shifman Vainshtein-Zakharov have proved to be an
extremely useful framework to learn about these masses[1-5]. However as the
quarks are not the physical states of QCD, there is no unique physical
definition of quark mass. The most widely used definition is the pole quark
mass $M_Q$, defined in perturbation theory as the position of the
singularity in the renormalized quark propagator. For the charm or bottom
quark, such definition becomes meaningless beyond perturbation theory.
Indeed it suffers from an intrinsic ambiguity generated by the so called
infrared renormalons[17]. The latter induce an additional uncertainties
which are presently estimated to be of the order of the QCD scale parameter
\allowbreak $\Lambda _{QCD}.$\\

On the other hand, the authors of ref.[1,2] observed that the radiative
corrections are too large in the case of the heavy two point correlator
associated to the vector current $\overline{Q}\gamma _\mu Q$, where Q
denotes the quantum field of the charm or bottom quark.\\

In spite of that, most QCD sum-rules determination of the c or b quark mass
have ignored the nonperturbative renormalons effects through the use the
pole quark mass definition. Moreover, they have performed QCD sum rules
analysis of the heavy correlator, with the perturbative contribution given
only up to the first order in the strong coupling constant.\footnote{Large order radiative corrections are still lacking, except a three loop
contributions to the heavy quark correlator, calculated in part numericaly
within the method of Pade approximants[20].}
\\
An alternative definition to the pole quark mass $M_Q$ is provided by the
Euclidean mass $m_Q^{Eucl}$ , which is defined in Landau gauge at the
Euclidean point $p^2=-M_Q^2$ . Although gauge dependent, the Euclidean mass
is free from renormalon ambiguities. It first has been introduced by SVZ [2]
in order to reduce the effects of higher multi-loop corrections on heavy
quarkonium correlators. Using relativistic Hilbert sum-rules, they have
obtained the following estimates: $m_c^{Eucl}=1.26\pm 0.02GeV$ and $%
m_{b=}^{Eucl}4.23\pm 0.05GeV.$ Subsequently; Reinders; Rubeinstein and
Yazaki [3] confirmed the predictions of SVZ, but they have advocated the use
of the Hilbert Moments at $Q^2=-q^2\neq 0$ for heavy quark systems. Later
on, Guberina et al. [4] and Reinders [5], following the same strategy, found
for the b Euclidean mass: $4.19\pm 0.06GeV$ and $4.17\pm 0.02GeV$
respectively. The discrepancies between these estimates come from two main
sources. First, the evolution of the experimental information on the
charmonium and the upsilon families. Indeed, only four resonances from six
in the bottomniun were observed in the beginning of eighties when these
papers appeared. Moreover, the measured hadronic parameters describing these
resonances ware contaminated with quite large experimental errors. The
second source of discrepancies is related to the theoretical inputs ($%
\langle \alpha _sG^2\rangle ,\alpha _{s,..}$) used in each work. Thus,
predictions of $m_b^{Eucl}$ are diverse and run from $4.13$ to $4.28GeV$,
while most predictions of $m_c^{Eucl}$ are gathered around the value $%
1.26GeV.$ Since, on the one hand, the present situation of the experimental
data on the $J/\Psi \quad and\quad \Upsilon $ systems has enormously
improved and, on the other hand, the main input parameters of QCD sum-rules
analysis are better under control than the past, a new and independent
determination of the Euclidean mass of the charm and bottom quarks is
certainly called for.\\

In the following, we will present an update of the theoretical analysis of
the vector channel of charmonium and bottomnium families based on
nonrelativistic Borel type sum-rules. We start from the expressions given in
the original work of Bell and Bertlman [6,7], which we expand , in the
infinite quark mass limit ($m_Q^{Eucl}\rightarrow \infty $), up to $\frac
1{m_Q^3}$ order. Next step will follow from the famous quark-hadron duality,
which will permit to confront two representations of the ratio of Borel
moments: one based on OPE calculations and the other on the data. An
important by-product of this analysis will be the prediction of the
Euclidean mass $m_Q^{Eucl}$.\\

The key input parameters most relevant to this sum-rules analysis are the
strong coupling constant $\alpha _s$ and the gluon condensate $\langle
\alpha _sG^2\rangle $. For the former, the world average central value $%
\alpha _s\left( M_z\right) =0.119$ is by now stable and the experimental
error taken by all measurements ranges from $0.003$ to 0.005 [8]. In this
work, our reference values will be: $\alpha _s=0.119\pm 0.005$. These values
will be combined to the formulae of the running strong coupling with two
loop accuracy, in order to evolve $\alpha _s$ down to the sum-rule energy
scale. As for $\langle \alpha _sG^2\rangle $, first estimated by SVZ from
the analysis of the charmonium sum-rules to be $0.038GeV^4[2]$, there has
been a lot of activity and discussion of its value. A number of independent
estimates have appeared in the literature [9]. Unlike the previous quark
mass determinations which use a fixed value of the gluon condensate, we will
take the conservative range of values:

\begin{equation}
\langle \alpha _sG^2\rangle =0.055\pm 0.025GeV^4
\end{equation}
\\
The plan of this paper is the following. Section 2 will be devoted to review
the most common mass definitions and their interrelations. In section 3 we
present the quarkonium correlation function and its non-relativistic Borel
moments. Special emphasize is put on their expansion in the infinite heavy
quark mass limit up to $\frac 1{M_Q^3}$order. In section 4, rigorous
analysis of sum-rules of the Euclidean mass of charm and bottom quarks is
performed and finally, our numerical estimates and discussion of the
uncertainties entailed by the sum-rules are given in section 5.

\section{Quark mass definitions}

The quark mass definition is quite confusing. Indeed, several definitions
have been proposed and the choice of one among them is often correlated to
the physical process of interest: each particular case requires its
definition. The most common ones rely on purely perturbative calculations.
The pole mass $M_Q$, the running mass in modified minimal subtraction scheme 
$\overline{m_Q}^{\overline{MS}}(\mu )$ and the Euclidean mass $m_Q^{Eucl}$
turn out to be very popular and are interrelated. Indeed, it is
straightforward to derive the following relations at-next-to-leading order
in QCD\ coupling constant [10],

\begin{equation}
\overline{m_Q}(\mu )=M_Q\{1-\frac 4{3\pi }\alpha _s+O(\alpha _s^2)\}
\end{equation}

\begin{equation}
m_Q^{Eucl}(\mu )=M_Q\{1-\frac{2\ln 2}\pi \alpha _s+O(\alpha _s^2)\}
\end{equation}
\\
Eq.(2) relates the pole quark mass to the running quark mass defined in the $%
\overline{MS}$ scheme and normalized at the scale $\mu ,$ while the
connection between the Euclidean quark mass in Landau gauge and the pole
mass is given by eq.(3).\\

The running coupling constant is given by,

\begin{equation}
\alpha _s(\mu )=\frac{4\pi }{\beta _0\ln (\frac{\mu ^2}{\Lambda _{QCD}^2})}%
\left\{ 1-\frac{\beta _1}{\beta _0^2}\frac{\ln \left[ \ln \left( \frac{\mu ^2%
}{\Lambda _{QCD}^2}\right) \right] }{\ln \left( \frac{\mu ^2}{\Lambda
_{QCD}^2}\right) }+...\right\}
\end{equation}
\\
which defines the QCD scale parameter $\Lambda _{QCD}$ at two loop
accuracy. $\beta _0$ and $\beta _1$ are the first beta functions governing
the evolution of $\alpha _s$. They are scheme independent and are
equal to [16],

\begin{equation}
\beta _0=11-\frac 23n_f,\ \qquad \beta _1=102-\frac{38}3n_f
\end{equation}
 \\
where $n_f$ is the number of flavors with mass below $\mu $. At
this stage, it should be understood that the number of flavors changes as $%
\mu $ crosses a quark threshold. As $\Lambda _{QCD}$ depends on $n_f$, value
of $\Lambda _{QCD}$ for different number of active flavors are defined by
keeping $\alpha _s$ continuous at the threshold scale $\mu =m_Q$. Therefore,
according to this prescription [11], we first extract the value of $\Lambda
_{QCD}^{(n_f=5)}$ from the present world average values of $\alpha
_s=0.119\pm 0.005$. The corresponding range is,

\begin{equation}
\Lambda _{QCD}^{(n_{f=}5)}=(235\pm _{60}^{75})MeV
\end{equation}
\\
Then owing to the matching condition $\alpha _s\left( \mu =m_b\right)
_{n_{f=4}}=\alpha _s\left( \mu =m_b\right) _{n_{f=5}},$ we relate $\Lambda
_{QCD}^{(n_f=4)}$ to $\Lambda _{QCD}^{(n_f=5)}$. The corresponding range of
values for $n_f=4$ is the following,

\begin{equation}
\Lambda _{QCD}^{(n_{f=}4)}=(340\pm _{80}^{95})MeV
\end{equation}
\\
Equations (6) and (7) constitute important inputs in this work.

\section{Quarkonium Sum-rules}
The basic amplitude considered in this sum-rules is the correlation function 
$\Pi _{\mu \upsilon }(q^2),$ induced by the electromagnetic vector current $%
j_\mu =\overline{Q}\gamma _\mu Q$ and defined by,

\begin{equation}
i\int d^4x\exp (iqx)\left\langle \Omega \mid j_\mu (x)j_\nu ^{+}(0)\mid
\Omega \right\rangle =\left( q_\mu q_\nu -q^2g_{\mu \nu }\right) \Pi (q^2)
\end{equation}
\\
The structure function $\Pi \left( q^2\right) $ is related to its imaginery
part through the dispersion relation:

\begin{equation}
\Pi (Q^2)=\frac 1\pi \int ds\frac{{Im}\Pi (s)}{(s+Q^2)}%
+subtraction,\qquad Q^2=-q^2
\end{equation}
\\
where ${Im}\Pi (s)$ is proportional to the total experimental
cross-section of $e^{+}e^{-}$ annihilition into final states with open or
hidden $Q$ flavores,

\begin{equation}
{Im} \Pi (s)=\frac s{16\pi ^2\alpha ^2e_Q^2}\sigma
(e^{+}e^{-}\rightarrow \overline{Q}Q)
\end{equation}
\\
$e_Q$ denotes the electric charge of the heavy quark Q.
\
In QCD sum-rules approach, $\Pi (Q^2)$ is calculated within Wilson Operator
Product Expansion (OPE)$[12],$ which take into account the perturbative
contributions as well as the non perturbative effects absorbed in the vacuum
matrix elements of operators such as $G^{\alpha \beta }G_{\alpha \beta }$ or 
$\overline{q}q$. In the case at hand, the leading and next-to-leading
perturbative contributions for the vector current can be found in
Schwinger's book [15],

\begin{equation}
{Im}\Pi (s)=\frac 1{8\pi }v(3-v^2)\{1+\frac 43\frac{\alpha _s}\pi [%
{\pi ^2}{2v^2}-\frac{(3+v)}4(\frac{\pi ^2}2-\frac 34)]\}\Theta (v^2)
\end{equation}
\\
where $v=(1-4\frac{m_Q^2}s)^{\frac 12}$, while the non-perturbative
corrections to $\Pi (s)$ will be represented by the leading contribution
which is proportional to the gluon consensate $\left\langle \alpha
_sG^2\right\rangle $ and is given by,

\begin{equation}
\Pi _{NP}(s)=\frac 1{48s^2}\{\frac{3\left( v^2+1\right) \left( 1-v^2\right)
^2}{2v^5}\ln (\frac{1+v}{1-v})-\frac{3v^4-2v^2+3}{v^4}\}<\alpha _sG^2>
\end{equation}
\\
In the sum-rules analysis, the subtracted dispersion relation in eq.(9) is
improved by the Borel transform $[2]$ leading to the exponential moments,

\begin{equation}
M(\sigma )=\frac 1\pi \int_{4m_Q^2}^\infty ds\exp (-s\sigma ){Im} \Pi
(s)
\end{equation}
\\
The exponential weight in (13) has the merit to cut off large contributions
and to take under control the effects of high dimension vacuum
condensates.This means an improvement of the convergence of the OPE as well
as a better enhancement of the quark-hadron duality.\\

In the present work, we shall be concerned with the non-relativistic version
of Borel sum-rules defined as,

\begin{equation}
M(\tau )=\frac 1\pi \int_0^\infty dE\exp (-E\tau ){Im}\Pi (E)
\end{equation}
\\
where we have introduced the new sum-rule parameter $\tau =4m_Q^{Eucl}\sigma 
$ and the heavy quark energy E, through the relation $s=$($2$m$_Q^{Eucl}+E$)$%
^2$. Moments present two important features: On the one hand, they can be
evaluated by using the detailed experimental data for ${Im}\Pi (s)(\sim
cross-section)$. On the other hand, the moments in eqs. (13) and (14) are
very sensitive to the heavy quark mass. Therefore, by equating the
theoretical moments based on OPE calculations to the corresponding ones that
use highly accurate experimental information in the heavy channel associated
to the vector current , one can make a reliable prediction of the heavy
quark mass. However, we will consider the ratio of the (Exponential / Borel)
moments instead of the moments themsleves: since in contrast to the moments,
the ratio has weaker dependance on higher order radiative corrections to the
unit operator of the OPE . As shown in ref.[7], the non-relativistic ratio
of Borel sum-rules is given by,

\begin{equation}
R_{NR}(\tau )=2m_Q^{Eucl}-\frac d{d\tau }\left[ \ln M_{NR}\left( \tau
\right) \right]
\end{equation}
\\
where at $\tau $-stability region the ground state mass is, in
principal provided by the minimum of $R_{NR}$.\\

The complete analytic expressions of the moments have been calculated by
Bell and Bertlmann in terms of Whittaker functions [6]. Owing to their
asymptotic properties, we expand them in powers of $\frac 1{m_Q^{Eucl}}$ up
to the next-to-next-to-leading order. The non-relativistic ratio is then
given by,

\[
R_{NR}(\tau )=2m_Q^{Eucl}\{1+\frac 34\omega [1-\frac 56\omega +\frac{10}%
3\omega ^2]-{\sqrt{\pi\omega} \over 3} {\alpha _s(\omega )}[1-(\frac 23+\frac 3{8\pi
^2})\omega +(\frac{107}{32}+\frac{51}{32\pi ^2})\omega ^2] 
\]
\begin{equation}
+\frac{2\ln 2}\pi \alpha _s(\omega )[1+\frac 54\omega ^2]+\frac{\pi ^2}3<%
\frac{\alpha _s}\pi G^2>\frac 1{(4m_Q^{Eucl2}\omega )^2}[1+\frac 43\omega
-\frac 5{12}\omega ^2]\}
\end{equation}
\\
where $\omega=\frac 1{m_Q^{Eucl}\tau}$.
Our expression in (16) agrees with the one given by ref.[13] which take into
account the mass corrections to the perturbative part of the ratio up
to-the-first order. However, we disagree with the expansion done in [14]. At
this stage some remarks are in order: The ratio $R_{NR}$ is a sign alternating
serie, therefore, it is not always fully justified to keep only the
next-to-leading mass corrections (NLO) and to ignore the
next-to-next-to-leading ones (NNLO). Indeed, for comparison, we show in
Table3 the results of the sum rules analysis for the c and b-Euclidean mass,
with the ratio (16) expanded up to the leading order, NLO and NNLO
respectively. As we see, in contrast to the case of the b-quark mass, the
effects of the higher order mass corrections are important for the c-quark
mass determination which is sensitive to the order of the $\frac
1{m_Q^{Eucl}}$ expansion.\\

On the other side, the experimental ratio can be thought of as a sum of zero
width resonances, plus a continuum contribution approximated by perturbation
theory above a threshold $s_T$. In this approximation $M(\sigma )$ is
evaluated through,

\begin{equation}
M_{\exp }(\sigma )=\frac 3{4\overline{d}^2e_Q^2}{\sum_{R} M_R}%
\Gamma _R\left( e^{+}e^{-}\rightarrow R\right) \exp \left( -\sigma
M_R^2\right) +\frac 1\pi \stackrel{\infty }{\int_{s_T}}ds\;e^{-\sigma s}%
{Im}\Pi _{Pert}\left( s\right)
\end{equation}
\\
where $\overline{\alpha }$ is the QED effective coupling constant.
\
Experimental data in $J/\Psi $ and $\Upsilon $ regions is very accurate. Six
resonances have been observed in the vector channel of charmonium and
bottomnium systems. Their masses and electronic widths are detailed in
the Review of Particle Properies[18]. They
come endowed with experimental errors $(\Delta M_R,\Delta \Gamma _R)$ that
we can safely ignor in our analysis. Indeed we show, for instance, that the
influence due to the variations of $\Gamma _R$ on the Euclidean mass
estimates is very small: it amounts to 0.1\% for the c-quark and to 0.2\%
for the b-quark as is indicated in Table 2.\\

As for the continuum threshold, we take it somewhere above the last
resonance of the family of interest. Then, we sensibly increase it in order
to test the stability of our results with respect to s$_T$. For our
analysis, the incertainties induced by the variations of $s_T$ from 4.5 to
8 $GeV^2$ for charmonium and from 11.2 to 15 $GeV^2$ for bottomnium, are
insignificant.

\section{Sum-rules analysis}

On the basis of the above discussion, we are now able to extract the
Euclidean mass of the charm and bottom quarks by confronting the two
representations shown in (16) and (17). Such confrontation should be
realized within the context of the duality between the hadron and
quark-gluon descriptions, invoked by SVZ. This means that the agreement
between the experiment and the theory should be ensured in a compromising
region of the Borel parameter $\tau $ \footnote{%
This is the so called fiducial window in SVZ' works}, over which both
contributions, perturbative and non-perturbative, to the theoretical ratio
of moments are on the same footing. In the other hand, the sum-rules defined
through the experimental ratio (17) should be saturated by the ground state
within the Borel region: The sum-rules become increasingly sensitive to the
ground state contribution, while the effects of the higher mass states are
less substantial and should not exceed a given moderate percentage of the
full contribution.\\

In Figure 1, we study the behavior of both the theoretical and the
experimental ratio as a function of the Borel scale $\sigma $ for the
Charmonium family.The input parameters are kept fixed to their central
values, $<\alpha _sG^2>=0.55GeV^4$ and $\Lambda _{QCD}^{(n_f=4)}=340MeV$. As
expected, we see that above $\sigma =0.65GeV^{-2}$ the sum-rule (17) is
dominated by the contribution of a single resonance $J/\Psi $ $.$ The
analysis of theoretical expression (16) for different values of $\sigma $
(or equivalently $\tau $) shows a (Borel) region of stability where a clear
minimum appears at $\sigma \sim 0.7GeV^{-2}$. The best fit of (16) to the
experimental curve is realized with $m_{c^{}}^{Eucl}=1.20GeV$. Next step, we
study the effects due to the variations of the inputs on the determination
of the Euclidean mass of Charm. With $<\alpha _sG^2>=0.6GeV^4,$ we show from
Figure 3 that the changes in $\Lambda _{QCD}^{(n_f=4)}$ within the range
given in (7) induce changes in $m_{c^{}}^{Eucl}$ from $1.176$ to $1.219GeV.$
Now, fixing the QCD scale parameter to its central value $340MeV$, the
variations of the gluon condensate in the interval $0.03-$ $0.08GeV^4$%
generate an uncertainties on the Charm quark mass of the order $\Delta
m_{c^{}}^{Eucl}=_{-29}^{+18}MeV$ (Table 1), which is necessary to the
restoration of the agreement between the two descriptions. Therefore, our
final results is,

\begin{equation}
m_c^{Eucl}=1.20\pm 0.034GeV
\end{equation}
\\
Similar analysis is performed for the Upsilon family. In figure 2, we plot
the curve of the theoretical ratio versus the experimental data. The
stability region shows up around $\sigma \simeq 0.2-0.24GeV^{-2}$ where the
best agreement between theory and the experiment is reached for $%
m_b^{Eucl}=4.18GeV$, with $\Lambda _{QCD}^{(n_f=5)}=235MeV$ and $<\alpha
_sG^2>=0.55GeV^4.$ The sensitivity of the b-quark Euclidean mass with respect to
the errors of the input quantities is displayed in figure 4 and Table 1,
from which we extract our final estimate:

\begin{equation}
m_b^{Eucl}=4.18\pm 0.037GeV
\end{equation}
\\
where all the errors have been added in quadrature.

\section{Conclusions}

We have extracted the numerical values of the charm and bottom quark
Euclidean masses from a rigorous analysis of the heavy correlator associated
to the vector current $\overline{Q}\gamma _\mu Q$, which we combine to the precise
experimental data on Charmonium and Upsilon systems. The analysis, performed
within the nonrelativistic Borel sum-rules, yields to the optimal results,

\[
m_c^{Eucl}=1.20\pm 0.034GeV
\]
\[
m_b^{Eucl}=4.18\pm 0.037GeV
\]
\\
From Table 2, we show the origins of the main uncertainties which enters the
errors on the c and b-quark masses. Besides the minor uncertainty coming from the
experimental data on $\Gamma _R$, they mainly result from the variations of
the gluon condensate in a conservative range of values and from the present
world average QCD coupling constant at $M_Z$ scale, evolved down to the sum
rules scale. Our predictions improve the previous determinations of the
Euclidean mass from Hilbert Quarkonium sum-rules[2-5]. Moreover, the
conversion of the values in (19) to the quark pole mass or $\overline{MS}$
mass of Charm and Beauty, through the relations (2) and (3) given to the
second order in $\alpha _s$ ,are in good agreement with the most recent
direct estimates[19]. Indeed, the corresponding ranges are respectively:

\begin{equation}
M_c^{}=1.49\pm 0.08GeV
\end{equation}
\begin{equation}
M_b=4.65\pm 0.06GeV
\end{equation}
and,
\begin{equation}
\overline{m}_c^{}(\overline{m}_c)=1.21\pm 0.08GeV
\end{equation}
\begin{equation}
\overline{m}_b^{}(\overline{m}_{b)}=4.20\pm 0.06GeV
\end{equation}
\\
where the errors on the pole masses and on $\overline{MS}$ running-masses
have increased, due to the additional uncertainty related to $\alpha _s(M_Q)$
variations in the formulae (2,3).

\section*{Acknowledgements}

M.C. is grateful to Abdus Salam Centre-ICTP for the warm hospitality and
financial support during his visit and would like to thank M.
Capdequi-Peyranere for valuable discussions.

\newpage
\listoffigures


        Fig. 1: Charm channel:Theoreical ratio (16)(continuous line) and the Experimental ratio
  (Dashed line) as a fuction of of the Borel sum-rule parameter $\sigma$,with
  $<\alpha_s G^2>=0.055 GeV^4$,  $\Lambda_{QCD}^{n_f=4}=235 MeV$ and $s_T=4.5GeV^2$.
\\

        Fig. 2: Similar to Fig.1 in the bottom channel, with
$\Lambda_{QCD}^{n_f=5} = 235 MeV$ and $s_T=11.2GeV^2$.
\\

        Fig. 3: Behaviour of the Charm Euclidean mass versus the QCD scale parameter,
 wit $<\alpha_s G^2> = 0.055 GeV^4$.
\\

Fig. 4: Similar to Fig. 3 for the bottom quark, with
$<\alpha_s G^2>=0.055GeV^4$.

\end{document}